\documentclass[showpacs,aps,prl,twocolumn,epsfig,superscriptaddress]{revtex4}
\usepackage{amssymb}
\usepackage{amsmath}
\usepackage{graphicx}

\begin{document}

\title{Self-consistent theory of phonon renormalization and electron-phonon
coupling near a 2D Kohn singularity}
\author{ O.V. Dolgov}
\affiliation{Max-Planck-Institut f\"{u}r Festk\"{o}rperforschung, Heisenbergstr.1, 70569
Stuttgart, Germany}
\author{O.K. Andersen}
\affiliation{Max-Planck-Institut f\"{u}r Festk\"{o}rperphysik, Heisenbergstr.1, 70569
Stuttgart, Germany}
\author{I.I. Mazin}
\affiliation{Center for Computational Materials Science, Naval Research Laboratory,
Washington, DC 20375, USA}

\begin{abstract}
We show that the usual expression for evaluating electron-phonon coupling
and the phonon linewidth in 2D metals with a cylindrical Fermi surface
cannot be applied near the wave vector corresponding to the Kohn
singularity. Instead, the Dyson equation for phonons has to be solved
self-consistently. If a self-consistent procedure is properly followed,
there is no divergency in either the coupling constant or the phonon
linewidth near the offending wave vectors, in contrast to the standard
expression.
\end{abstract}

\date{\today }
\maketitle

First principles calculations of the phonon spectra and electron-phonon
coupling in MgB$_{2}$ (see Ref. \cite{MA} for a review) have determined that
the interaction mainly responsible for superconductivity in this material
is coupling of small-$q$, high-energy optical phonons of a particular symmetry with
approximately parabolic nearly two-dimensional hole bands forming
practically perfect circular cylinders occupying only a small fraction of
the Brillouin zone. It was realized early enough\cite{kong} that the case of
an \textit{ideal} 2D cylinder leads to a divergency in the calculated phonon
linewidth at the 2D Kohn singularity, $q=2k_{F},$ which presents serious
difficulties in calculating the electron-phonon coupling function. One
option that was exploited was to use an analytical integration for the wave
vectors comparable with, or smaller than $2k_{F}$\cite{kong,amy}, and a
numerical one for the larger vectors.

It was also pointed out\cite{MA,an} that this singularity gets stronger when
the Fermi surface gets smaller, while the integrated electron-phonon
coupling (for 2D parabolic bands) does not change. While a perfectly
cylindrical Fermi surface is an idealized construction, deviations may be
quite small, it seems, on the first glance, unphysical that all phonons with 
$|\mathbf{q}|=2k_{F}$ have infinite linewidth. Note that the problem is
not specific for MgB$_{2}:$ it occurs for any 2D material sporting Kohn
singularities. In particular, the hypothetical hexagonal LiB, a subject of a
substantial recent interest, has $\sigma $ bands that are even more 2D that
those in MgB$_{2}$ \cite{lib} and the described problem is even more
pronounced.

This intuition is correct. In this paper we show that close to a Kohn
anomaly standard formulas for calculating electron-phonon interaction (EPI)
become incorrect, and new, self-consistent expressions replace them. These
expressions have no singularities, and exhibit a much more natural,
reasonably smooth, \textbf{q}-dependence of the phonon self-energy.

To start with, we shall remind the readers the standard formalism. We first
define the retarded phonon Green function%
\begin{equation*}
D^{\alpha \beta }\left( \mathbf{q},t\right) \equiv -i\theta (t)\left\langle %
\left[ u_{\mathbf{q}}^{\alpha }(t)u_{-\mathbf{q}}^{\beta }(0)\right]
\right\rangle ,
\end{equation*}%
where $\left[ ...\right] $ is a commutator and $\left\langle
...\right\rangle $ denotes statistical averaging. The displacement operator $%
u_{\mathbf{q}}^{\alpha }$\ in the $\alpha -$direction can be expressed via
the phonon eigenvectors $e_{\mathbf{q}\nu }$\ and frequencies squared $%
\omega _{\mathbf{q}\nu }^{2}$.
\begin{equation}
\mathbf{u}_{\mathbf{q}}=\sum_{\nu }\left( \frac{1}{2M\omega _{\mathbf{q}\nu }%
}\right) \mathbf{e}_{\mathbf{q}\nu }(a_{\mathbf{q}\nu }+a_{-\mathbf{q}\nu
}^{\dag }).  \label{u}
\end{equation}%

{For simplicity, a primitive lattice with a single kind of ions
with a single mass }$M$ {will be considered below. Also, atomic 
(Hartree) units will be used throughout the paper. In this case the 
``bare'' phonon Green function has a form} 
\begin{equation}
D_{0}(\mathbf{q},\omega )=\frac{1}{2M\omega _{0}}\left[ \frac{1}{\omega
-\omega _{0}+i\delta }-\frac{1}{\omega +\omega _{0}+i\delta }\right] ,
\label{D0}
\end{equation}%
where $\omega _{0}$ is the bare phonon frequency, before accounting for
electron phonon coupling (screening by electrons). Without losing
generality, it can be assumed to be $\mathbf{q}$ independent.
Correspondingly, the full Green function is

\begin{equation}
D(\mathbf{q},\omega )=\frac{1}{2M\omega _{\mathbf{q}}}\left[ \frac{1}{\omega
-\omega _{\mathbf{q}}+i\Gamma _{\mathbf{q}}}-\frac{1}{\omega +\omega _{%
\mathbf{q}}+i\Gamma _{\mathbf{q}}}\right] ,  \label{D}
\end{equation}%
where $\omega _{\mathbf{q}}$ is the renormalized (observable) frequency, and 
$\Gamma _{\mathbf{q}}$ is damping (phonon linewidth) due to EPI \cite%
{footnote}.

The Dyson equation reads%
\begin{equation}
D^{-1}(\mathbf{q},\omega )=D_{0}^{-1}(\mathbf{q},\omega )-\Pi (\mathbf{q}%
,\omega ),  \label{Dyson}
\end{equation}%
where the polarization operator in the lowest approximation (as usual, the
Migdal theorem\cite{migdal} allows neglecting the vertex corrections) along
the real frequency axis at $T=0$ has a form ($a$ is the lattice constant in
the plane)%
\begin{widetext}
\begin{equation}
\Pi (\mathbf{q},\omega )=-2i\int \left\vert g_{\mathbf{k,k+q}%
}^{0}\right\vert ^{2} G_{0}(\mathbf{k+\frac q 2},\varepsilon +\frac \omega
2)G_{0}(\mathbf{k-\frac q 2},\varepsilon -\frac \omega 2) \frac{a^2d^{2}k}{%
(2\pi )^{2}}\frac{d\varepsilon }{2\pi }.  \label{Pi}
\end{equation}%
Here $g_{\mathbf{k,k+q}}^{0}$ is the \textit{bare} electron-ion scattering
 matrix element (the commonly used EPI matrix element differs in that the 
potential gradient is replaced by the derivatives with respect to the normal 
phonon coordinates)%
\begin{equation}
g_{\mathbf{k,k+q}}^{0}=\frac{1}{a^{2}}\int_{a^{2}}\psi _{\mathbf{k+q}}^{\ast
}(\mathbf{r})\nabla V(\mathbf{r})\psi _{\mathbf{k}}(\mathbf{r})d^{2}r
\label{matr}
\end{equation}%
where 
\begin{equation}
G_{0}(\mathbf{k},\varepsilon )=\frac{1}{\varepsilon -\varepsilon _{0}(%
\mathbf{k})+i\delta \mathrm{sign}(k-k_{F})}  \label{G0}
\end{equation}%
is the bare electron Green function, and $\varepsilon _{0}(\mathbf{k}%
)=(k_{x}^{2}+k_{y}^{2}-k_{F}^{2})/2m$. The \textit{renormalized }phonon
frequency and the phonon \textit{line width} $\Gamma _{\mathbf{q}}$ are
determined by the \textit{pole} of \ the phonon Green function $D(\mathbf{q}%
,\omega )$ or 
\[
D^{-1}(\mathbf{q},\omega _{\mathbf{q}}+i\Gamma _{\mathbf{q}})\equiv
D_{0}^{-1}(\mathbf{q},\omega _{\mathbf{q}}+i\Gamma _{\mathbf{q}})-\Pi (%
\mathbf{q},\omega _{\mathbf{q}}+i\Gamma _{\mathbf{q}})=0. 
\]%
This leads to%
\begin{equation}
\omega _{\mathbf{q}}^{2}=\omega _{0}^{2}+\frac{1}{M}\mathrm{Re}\Pi (\mathbf{q%
},\omega _{\mathbf{q}}+i\Gamma _{\mathbf{q}})+\Gamma _{\mathbf{q}}^{2},
\label{omD}
\end{equation}%
and

\begin{equation}
\Gamma _{\mathbf{q}}=-\frac{1}{2M\omega _{\mathbf{q}}}\mathrm{Im}\Pi (%
\mathbf{q},\omega _{\mathbf{q}}+i\Gamma _{\mathbf{q}}).  \label{gm}
\end{equation}

The next standard step, following Ref. \cite{allen}, is to expand the
polarization operator%
\begin{equation}
\Pi (\mathbf{q},\omega +i\delta )=-2\sum_{\mathbf{k}}\left\vert g_{\mathbf{%
k,k+q}}^{0}\right\vert ^{2}\frac{n_{\mathbf{k}}-n_{\mathbf{k+q}}}{%
\varepsilon _{0}(\mathbf{k+q})-\varepsilon _{0}(\mathbf{k})-\omega -i\delta }
\label{pi1}
\end{equation}%
to first order in frequency just above the real axis of the
complex frequency $\omega .$ In this case 
\begin{equation}
\omega _{app\mathbf{q}}^{2}=\frac{1}{M}\Pi (\mathbf{q,}0)\left( 1+O(\omega
_{0}^{2}/\varepsilon _{F}^{2})\right)   \label{om1}
\end{equation}%
and%
\begin{equation}
\Gamma _{\mathbf{q}}^{app}=-\frac{1}{2M}\left. \frac{d\mathrm{Im}\Pi (%
\mathbf{q},\omega )}{d\omega }\right\vert _{\omega =0}\equiv \frac{\pi }{M}%
\sum_{\mathbf{k}}\left\vert g_{\mathbf{k,k+q}}^{0}\right\vert ^{2}\delta
\lbrack \varepsilon _{0}(\mathbf{k+q})]\delta \lbrack \varepsilon _{0}(%
\mathbf{k})]  \label{g1}
\end{equation}%
where the factor $\left( n_{\mathbf{k}}-n_{\mathbf{k+q}}\right) \delta
\lbrack \varepsilon _{0}(\mathbf{k+q})-\varepsilon _{0}(\mathbf{k})-\omega ]$
has been changed to $-\omega \delta \lbrack \varepsilon _{0}(\mathbf{k+q}%
)]\delta \lbrack \varepsilon _{0}(\mathbf{k})]$. According to Ref.\cite%
{allen2} \textquotedblleft Except for extremely pathological energy bands,
it is an excellent approximation\textquotedblright . Unfortunately, MgB$_{2}$
and some other recently discovered superconductors are examples where the
energy bands are, in some aspects, pathological. Formally the expression (%
\ref{g1}) for 2D-system is divergent near a Kohn anomaly $q\rightarrow 2k_{F}
$, and we have $\Gamma _{\mathbf{q}}\geq $ $\omega _{\mathbf{q}}$, i.e.
phonons are not well defined quasiparticles. To describe electron-phonon
interaction in these systems we have to calculate the polarization operator $%
\Pi (\mathbf{q},Z)$ for a complex frequency $Z$ and solve Eqs. (\ref{omD},\ref%
{gm}).

\section{Complex polarization operator}

Let us consider a model with a cylindrical Fermi surface of radius $%
k_{F}$ whose electrons interact with an optical phonon with a bare
frequency $\omega _{0},$ and a momentum-independent
matrix element $g^{0}$  (we also neglect
possible warping of the Fermi-surface cylinder, $cf.$ Ref.\cite{cal-maur}).
In this case the imaginary part of Eq. \ref{pi1} reads%
\[
\mathrm{Im}\Pi (\mathbf{q},\omega +i\delta )=-2\pi \left\vert
g^{0}\right\vert ^{2}\sum_{\mathbf{k}}\left\{ \theta \left[ \varepsilon _{0}(%
\mathbf{k+q})\right] -\theta \left[ \varepsilon _{0}(\mathbf{k})\right]
\right\} \delta \left[ \varepsilon _{0}(\mathbf{k+q})-\varepsilon _{0}(%
\mathbf{k})-\omega \right] 
\]%
or 
\begin{equation}
\mathrm{Im}\Pi (\mathbf{q},\omega )=-\frac{m\left\vert g^{0}\right\vert ^{2}%
}{2\pi a^{2}q}\int_{\max \{k_{F},m\omega /q+q/2\}}^{\sqrt{k_{F}^{2}+2m\omega 
}}\mathrm{Re}\frac{dk}{\sqrt{1-\left( q/2k+m\omega /kq\right) ^{2}}}=-\frac{%
m\left\vert g^{0}\right\vert ^{2}}{\pi a^{2}Q}\left[ \mathrm{Re}\sqrt{%
1-(Q-\Omega)^{2}}-\mathrm{Re}\sqrt{1-(Q+\Omega)^{2}}\right] ,
\end{equation}%
where $Q=q/2k_{F},$ and $\Omega=\omega /qv_{F}$ ($v_F$ is the Fermi
velocity),

To find the polarization operator for a complex frequency $Z=\omega+i \gamma$, we use the
Hilbert transformation%
\[
\Pi (\mathbf{q},Z)=\frac{1}{\pi }P\int_{-\infty }^{\infty }\frac{dE}{E-Z}%
\mathrm{Im}\Pi (\mathbf{q},E). 
\]%
The result is%
\begin{equation}
\Pi (\mathbf{q},Z)=-\frac{m\left\vert g^{0}\right\vert ^{2}}{\pi a^{2}Q}%
\int_{C}dxdy\frac{x}{-(Z/4\varepsilon _{F})^{2}/Q^{2}+x^{2}},  \label{p_comp}
\end{equation}%
where $\varepsilon _{F}=k_{F}^{2}/2m$, $x=k_{x}/k_{F}$, $y=k_{y}/k_{F}$,
and $N(0)=m/2\pi a^{2}$ is the density of states at the Fermi level, per
spin. The integration is performed over the range $(x-Q)^{2}+y^{2}<1$. The
substitution $x-Q=r\cos \varphi ,$ $y=r\sin \varphi $ leads to%
\[
\int_{C}dxdy\Rightarrow \int_{0}^{1}rdr\int_{0}^{2\pi }d\varphi . 
\]%
\begin{equation}
\Pi (\mathbf{q},Z)=-\frac{m\left\vert g^{0}\right\vert ^{2}}{2\pi a^{2}}%
\left[ 2-(1-\frac{Z}{4\varepsilon _{F}Q^{2}})\sqrt{1-\left( Q-\frac{Z}{%
4\varepsilon _{F}Q}\right) ^{-2}}-\{Z\rightarrow -Z\}\right] .\label{p_exact}
\end{equation}
{ The branches of the
square roots are chosen so as to get the correct behavior }$\Pi \propto
q^{2}V_{F}^{2}/\omega ^{2}${at large frequencies }$\omega$ ($\omega ={\rm Re}%
Z$).
For a 2D system for $\Pi (\mathbf{q},\omega )$ on the real frequency axis
one can write\cite{stern,ando,zhang,IT}:%
\begin{eqnarray}
\mathrm{Re}\Pi (\mathbf{q},\omega +i\delta ) &=&-\frac{m\left\vert
g^{0}\right\vert ^{2}}{2\pi a^{2}Q}\left[ 2Q-(Q-\Omega)\mathrm{Re}\sqrt{%
1-(Q-\Omega)^{-2}}-(Q+\Omega)\mathrm{Re}\sqrt{1-(Q+\Omega)^{-2}}\right]  \label{PiR} \\
\mathrm{Im}\Pi (\mathbf{q},\omega +i\delta ) &=&-\frac{m\left\vert
g^{0}\right\vert ^{2}}{2\pi a^{2}Q}\left[ \mathrm{Re}\sqrt{1-(Q-\Omega)^{2}}-%
\mathrm{Re}\sqrt{1-(Q+\Omega)^{2}}\right] .  \nonumber
\end{eqnarray}%

First, we see that the imaginary part a finite for all values of the
wavevector $q$ and vanish inside the Landau-damping cone $q<\omega /v_{F}$
(more exactly, at $Q<\omega /4\varepsilon _{F}-(\omega /4\varepsilon
_{F})^{2}).$ It has two maxima: one is rather small, $\mathrm{Im}\Pi (\omega
)\simeq -$ $\frac{m\left\vert g^{0}\right\vert ^{2}\sqrt{\omega
/4\varepsilon _{F}}}{\pi a^{2}},$ at $Q\simeq 1-\omega /4\varepsilon _{F},$
while the other has an antiadiabatical behavior $\mathrm{Im}\Pi (\omega
)\simeq -$ $\frac{m\left\vert g^{0}\right\vert ^{2}\sqrt{2\varepsilon
_{F}/\omega }}{\pi a^{2}}$ and occurs at a very low frequency $Q\simeq
\omega /4\varepsilon _{F}+(\omega /4\varepsilon _{F})^{2}.$

But if we expand the polarization operator at small frequencies we recover a
standard result (see, \textit{e.g}. Ref.\cite{kong,an}) 
\begin{equation}
\Pi _{app}(\mathbf{q},\omega +i\delta )\simeq -\frac{m\left\vert
g^{0}\right\vert ^{2}}{\pi a^{2}}\left[ 1+\frac{i\omega /4\varepsilon _{F}}{Q%
\sqrt{1-Q^{2}}}\theta (1-Q)-\sqrt{1-1/Q^{2}}\theta (Q-1)\right] ,
\label{PiA}
\end{equation}%
where the imaginary part diverges at $q\rightarrow 0$ and $q\rightarrow
2k_{F}$. The real part of Eq. \ref{PiR} practically coincides with the real
part of Eq. \ref{PiA} except inside the Landau damping region.

At $\omega \rightarrow 0$ and finite $\mathbf{q}$ we get 
\begin{equation}
\mathrm{Im}\Pi _{app}(\mathbf{q},\omega ) \simeq 
-\frac{\omega }{4\varepsilon _{F}}\frac{m\left\vert g^{0}\right\vert ^{2}%
}{\pi a^{2}}\frac{\theta (1-Q)}{Q\sqrt{1-Q^{2}}}.   \label{im_small} 
\end{equation}

In the opposite limit $\mathrm{Im}\Pi (\mathbf{q},\omega )\equiv 0$ for $%
q\leqslant m\omega /\sqrt{k_{F}^{2}-m\omega }$. For the real part one can
set $\omega =0$ in Eq. \ref{PiA}:

\[
\mathrm{Re}\Pi (\mathbf{q},0)=-2\left\vert g^{0}\right\vert ^{2}\sum_{%
\mathbf{k}}\theta (\left\vert \mathbf{k}\right\vert -k_{F})\theta
(k_{F}-\left\vert \mathbf{k+q}\right\vert )\frac{1}{\varepsilon _{0}(\mathbf{%
k+q})-\varepsilon _{0}(\mathbf{k)}}. 
\]%
In this case%
\begin{equation}
\mathrm{Re}\Pi (\mathbf{q},0)\simeq -\frac{m\left\vert g^{0}\right\vert ^{2}%
}{\pi a^{2}}\left[ 1-\theta (q-2k_{F})\sqrt{1-(2k_{F}/q)^{2}}\right] =-\frac{%
m\left\vert g^{0}\right\vert ^{2}}{\pi a^{2}}\left[ 1-\theta (Q-1)\sqrt{%
1-Q^{-2}}\right] .  \label{re_small}
\end{equation}%
\textbf{\ }In the opposite limit 
\[
\mathrm{Re}\Pi (\mathbf{q}\rightarrow 0,\omega )\approx -\frac{m\left\vert
g^{0}\right\vert ^{2}}{2\pi a^{2}}\left( \frac{qk_{F}}{m\omega }\right)
^{2}. 
\]

The momentum dependence of the absolute values of the imaginary (the upper
panel) and the real parts (the lower panel) of Eqs. \ref{PiR} (solid lines)
and \ref{PiA} (short-dash lines) at $\omega =\omega _{0}=90$ meV as the
functions of the reduced wavevector $q/k_{BZ}$ ($k_{BZ}=\pi/a$ is the
Brillouin zone vector, or the radius of the Wigner-Seitz cylinder) are shown
in Fig. 1. Fermi vectors $k_F/k_{BZ}=0.075,k_F/k_{BZ}=0.1,k_F/k_{BZ}=0.15, $%
and $k_F/k_{BZ}=0.2$ correspond to $\varepsilon _{F}$ =0.15 eV, 0.27 eV,
0.60 eV, and 1.07 eV, respectively.

Along the imaginary (Matsubara) axis the polarization operator has the
following form%
\begin{equation}
\Pi _{M}(\mathbf{q},i\omega _{n})=-\frac{m\left\vert g^{0}\right\vert ^{2}}{%
\pi a^{2}}\left[ 1+\frac{\sqrt{Q^{4}-Q^{2}-\omega _{n}/4Q\varepsilon _{F}+%
\sqrt{(Q^{4}-Q^{2}-\omega _{n}/4Q\varepsilon _{F})^{2}+(Q\omega
_{n}/4\varepsilon _{F})^{2}}}}{\sqrt{2}Q^{2}}\right] ,  \label{mats}
\end{equation}%
where $\omega _{n}=2\pi nT.$ $T$ is temperature, $n$=$0,\pm 1,\pm 2,...\pm
\infty $. $\Pi _{M}(\mathbf{q},i\omega _{n}\rightarrow 0)$ coincides with
the Eq. \ref{re_small}.
\end{widetext}
\begin{figure}[htbp]
\centering \includegraphics[angle=0,width=0.99\linewidth]{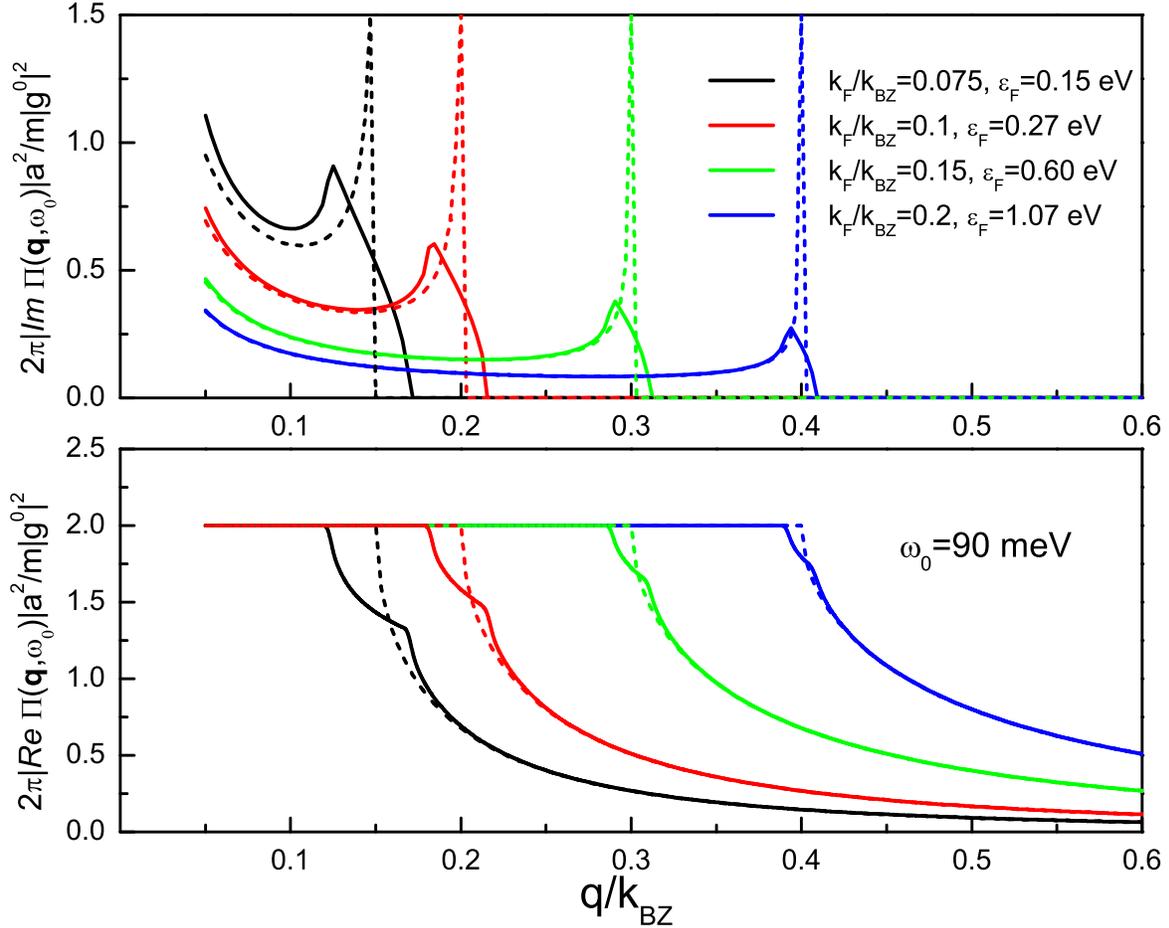}
\caption{(color online) The imaginary and real parts of the normalized polarization
operator $2\protect\pi \left\vert \Pi (\mathbf{q},\protect\omega =\protect%
\omega _{0})\right\vert a^{2}/m\left\vert g^{0}\right\vert ^{2}$ as a function of the
reduced wavevector $q/k_{BZ}$, for different fillings $k_{F}/k_{BZ}$.
Solid lines represent
the exact results, and the dashed lines the approximate solution (Eq. 
\ref{PiA}). The four different sets correspond, from left to right, to
four increasing values of $k_F/k_{BZ}$.
}
\label{bands}
\end{figure}

\section{Phonon renormalization in 2D systems}

First let us consider the \textit{approximate} polarization operator from
Eq. \ref{PiA}. Then%
\begin{equation}
\omega _{\mathbf{q}}^{app}=\omega _{0}\sqrt{1-2\zeta \left[ 1-\theta
(q-2k_{F})\sqrt{1-(2k_{F}/q)^{2}}\right] }  \label{om_small}
\end{equation}%
and according to Eqs. \ref{g1} and \ref{PiA} 
\begin{equation}
\Gamma _{\mathbf{q}}^{app}=\frac{m\left\vert g^{0}\right\vert ^{2}}{2\pi
a^{2}MV_{F}q}\frac{\theta (2k_{F}-q)}{\sqrt{1-(q/2k_{F})^{2}}}=\frac{%
\zeta \omega _{0}^{2}}{4\varepsilon _{F}}\frac{\theta (1-Q)}{Q\sqrt{%
1-Q^{2}}},  \label{gam_small}
\end{equation}%
where we have introduced, following Ref. \cite{froeh}, an \textit{auxiliary}
coupling constant $\zeta =N(0)\left\vert g^{0}\right\vert ^{2}/M\omega
_{0}^{2}$ (some authors use another dimensionless constant $\lambda
_{0}=2\zeta $). $\Omega _{\mathbf{q}}=\omega _{\mathbf{q}}^{app}+i\Gamma _{%
\mathbf{q}}^{app}$ gives the renormalized frequency and the damping (see
Eqs. \ref{om1},\ref{g1}). For $q\leq 2k_{F}$%
\begin{equation}
\frac{\Gamma _{\mathbf{q}}^{app}}{\omega _{\mathbf{q}}^{app}}=\frac{%
\varsigma \omega _{0}}{4\varepsilon _{F}\sqrt{1-2\zeta }}\frac{1}{Q\sqrt{%
1-Q^{2}}}.  \label{r_app}
\end{equation}%
This expression diverges in the limits $Q\rightarrow 0$\ and $Q\rightarrow
1. $

Turning now to the exact Eq. \ref{p_exact}, we observe that in the small $q$
limit the polarization operator becomes real:%
\begin{equation}
\Pi (\mathbf{q}\rightarrow 0,\omega )=-\frac{m\left| g^{0}\right| ^{2}}{2\pi
a^{2}}\left( \frac{4\varepsilon _{F}}{\omega }\right) ^{2}Q^{2}.  \label{p0}
\end{equation}%
Solving the equation%
\begin{equation*}
D^{-1}(\mathbf{q\rightarrow }0,\omega )=\omega ^{2}-\omega _{0}^{2}+\zeta
\omega _{0}^{2}\left( \frac{4\varepsilon _{F}}{\omega }\right) ^{2}Q^{2}=0
\end{equation*}%
one gets%
\begin{equation*}
\omega _{\mathbf{q\rightarrow }0}=\sqrt{\sqrt{\frac{\omega _{0}^{4}}{4}%
-16\zeta \omega _{0}^{2}\varepsilon _{F}^{2}Q^{2}}+\frac{\omega _{0}^{2}}{2}}%
.
\end{equation*}%
We choose the branch of the complex square root that gives $\omega _{\mathbf{%
q\rightarrow }0}\rightarrow \omega _{0}$ for $\zeta \rightarrow 0.$ This
means that up to $O(Q^{2})$ there is no damping for the phonon. The phonon
spectral function $F(\omega )=-\frac{1}{\pi }{\rm Im}D(\mathbf{q}%
\rightarrow 0,\omega )$ shows a narrow peak at this frequency and $\Gamma _{%
\mathbf{q\rightarrow }0}/\omega _{\mathbf{q\rightarrow }0}$ vanishes. The
situation at $Q\rightarrow 1$ is similar. In the lowest order in $\omega
/\varepsilon _{F}$ we have%
\begin{equation*}
\Pi (q\mathbf{\rightarrow }2k_{F},\omega )=-\frac{m\left| g^{0}\right| ^{2}}{%
\pi a^{2}}\left[ 1-(1+i)\sqrt{\omega /8\varepsilon _{F}}\right] 
\end{equation*}%
This leads to 
\begin{eqnarray*}
\omega _{2k_{F}} &\simeq &\omega _{0}\sqrt{1-2\zeta }, \\
\Gamma _{2k_{F}} &=&\frac{\zeta \omega _{0}^{2}}{\omega _{2k_{F}}}\sqrt{%
\frac{\omega _{2k_{F}}}{8\varepsilon _{F}}},
\end{eqnarray*}%
and%
\begin{equation}
\frac{\Gamma _{2k_{F}}}{\omega _{2k_{F}}}=\frac{\zeta }{1-2\zeta }\sqrt{%
\frac{\omega _{0}\sqrt{1-2\zeta }}{8\varepsilon _{F}}}  \label{rat}
\end{equation}%
This ratio remains finite in the limit $Q\rightarrow 1,$ although the
approximate expression of Eq. \ref{r_app} diverges for any system with a
cylindrical Fermi surface . The main point is that in both cases the well
known popular formula%
\begin{eqnarray}
\frac{\Gamma _{\mathbf{q}}}{\omega _{\mathbf{q}}} &=&\frac{\left|
g_{0}^{2}\right| }{M}\sum_{\mathbf{k}}\delta (\varepsilon _{\mathbf{k}%
})\delta (\varepsilon _{\mathbf{k+q}})  \label{g/w} \\
&=&\frac{\left| g_{0}^{2}\right| }{M(2\pi a)^{2}}\int \frac{dk}{\left| 
\mathbf{v}(\mathbf{k})\times \mathbf{v}(\mathbf{k}+\mathbf{q})\right| } 
\notag
\end{eqnarray}%
is not valid near the Kohn singularity.

The results for the linewidth $\Gamma _{\mathbf{q}}$ and the renormalized
phonon frequency $\omega _{\mathbf{q}}$ obtained by using the approximate
polarization operator as functions of the reduced wavevector $Q=q/2k_{F}$
are shown in Fig.2 by red lines. Parameters are following : the bare phonon
frequency $\omega _{0}=90$ meV , the bare constant of EPI $\ \zeta =1/4$.The
ratio $k_{F}/k_{BZ}$ is equal to 0.17. It corresponds to $\varepsilon
_{F}=0.2$ eV.

\begin{figure}[tbph]
\centering \includegraphics[angle=0,width=0.99\linewidth]{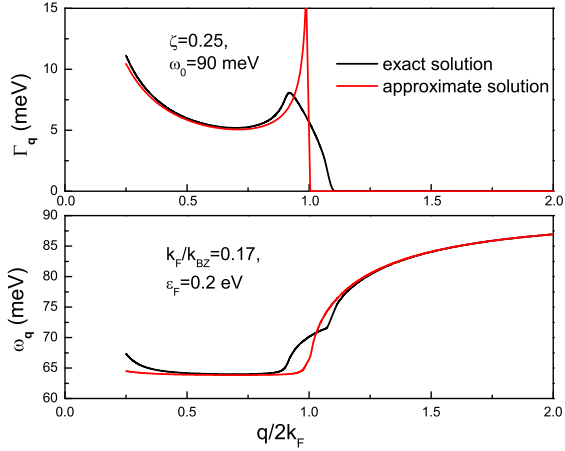}
\caption{(color online) The linewidth $\Gamma _{\mathbf{q}}$ and the renormalized phonon
frequency $\protect\omega _{\mathbf{q}}$ obtained by using the
approximation of Eq.
\ref{PiA} and the exact expression Eq. \ref{p_exact}, for
the following parameters : the bare phonon frequency $\protect\omega %
_{0}=90$ meV , the bare constant of EPI $\ \protect\zeta =1/4$. The filling
corresponds to $k_{F}/k_{BZ}=0.17$ and $\protect\varepsilon _{F}=0.2$ eV.}
\end{figure}

The approximate result agrees with that of Ref.\cite{an} (their Fig.4). The
exact result (black lines) has been obtained by a numerical solution of Eqs. %
\ref{omD},\ref{gm} using the polarization operator from the Eq. \ref{p_exact}%
. The latter, in contrast to the approximate expression, shows two shoulders
in the wavevector dependence of the renormalized frequency $\omega _{\mathbf{%
q}}$ (see also $\left\vert \mathrm{Re}\Pi (\mathbf{q},\omega
_{0})\right\vert $ in the bottom panel of Fig. 2). The first one corresponds
to the maximum of $\left\vert \mathrm{Im}\Pi (\mathbf{q},\omega )\right\vert 
$\ (or $\Gamma _{\mathbf{q}}$) and the second one to vanishing of these
values.

\section{Electron self-energy}

The electron self-energy is expressed via the electron and phonon Green
functions:%
\begin{equation*}
\Sigma (p)=ig^{0}\int G(p-k)D(k)\Gamma (p,p-k,k)\frac{d^{D+1}k}{(2\pi )^{D+1}%
},
\end{equation*}%
where $p=\{\mathbf{p},\varepsilon \}$. It was shown by Migdal \cite{migdal}
that one can neglect the vertex corrections $\Gamma (p,p-k,k)\simeq g^{0}(1+%
\sqrt{m/M})$ and that the function $G(\varepsilon ,\mathbf{k})$ differs from
the bare electron Green function (Eq. \ref{G0}) only in the narrow interval
of momenta $\left| k-k_{F}\right| \lesssim \omega _{ph}/V_{F}$ and
frequencies $\left| \omega \right| \lesssim \omega _{ph}.$ Thus, the full
electron Green function $G(p)$ can be substituted by the corresponding
function for noninteracting electrons (Eq. \ref{G0}). Using the Eq. \ref{D},
the electron self-energy for $T=0$ becomes (see, \textit{e.g.}, Refs.\cite%
{mahan}, \cite{am})%
\begin{widetext}
\begin{equation*}
\Sigma (\mathbf{k},\omega )=\sum_{\mathbf{q}}\delta (\varepsilon _{\mathbf{%
k+q}})\frac{\left\vert g^{0}\right\vert ^{2}}{2M\omega _{\mathbf{q}}}\int
d\xi \left[ \frac{\theta (\xi )}{\omega -\xi -\omega _{\mathbf{q}}+i\Gamma _{%
\mathbf{q}}}+\frac{\theta (-\xi )}{\omega -\xi +\omega _{\mathbf{q}}-i\Gamma
_{\mathbf{q}}}\right] . 
\end{equation*}%
This is a trivial generalization of the standard expressions onto the finite
phonon linewidth case. Let us average the self energy over the Fermi surface

\begin{eqnarray*}
\Sigma (\omega ) &=&\frac{1}{N(0)}\sum_{\mathbf{k}}\delta (\varepsilon _{%
\mathbf{k}})\Sigma (\mathbf{k},\omega ) \\
&=&-\frac{1}{N(0)M}\sum_{\mathbf{k}}\sum_{\mathbf{q}}\delta (\varepsilon _{%
\mathbf{k}})\delta (\varepsilon _{\mathbf{k+q}})\frac{\left\vert
g^{0}\right\vert ^{2}}{4\omega _{\mathbf{q}}}\left\{ \ln \frac{\Gamma _{%
\mathbf{q}}^{2}+(\omega _{\mathbf{q}}-\omega )^{2}}{\Gamma _{\mathbf{q}%
}^{2}+(\omega _{\mathbf{q}}+\omega )^{2}}+2i\left[ \tan ^{-1}\left( \frac{%
\omega _{\mathbf{q}}-\omega }{\Gamma _{\mathbf{q}}}\right) -\tan ^{-1}\left( 
\frac{\omega _{\mathbf{q}}+\omega }{\Gamma _{\mathbf{q}}}\right) \right]
\right\} .
\end{eqnarray*}%
The limit $\lambda =-\lim_{\omega \rightarrow 0}\mathrm{Re}\Sigma (\omega
)/\omega $ is nothing but the standard electron-phonon coupling constant%
\begin{equation}
\lambda _{\Gamma }=\frac{1}{MN(0)}\sum_{\mathbf{k}}\sum_{\mathbf{q}}\delta
(\varepsilon _{\mathbf{k}})\left\vert g^{0}\right\vert ^{2}\delta
(\varepsilon _{\mathbf{k+q}})\frac{1}{\Gamma _{\mathbf{q}}^{2}+\omega _{%
\mathbf{q}}^{2}}=\sum_{\mathbf{q}}\tilde{N}_{\mathbf{q}}(0)\frac{\left\vert
g^{0}\right\vert ^{2}}{M}\frac{1}{\Gamma _{\mathbf{q}}^{2}+\omega _{\mathbf{q%
}}^{2}},  \label{lamg}
\end{equation}%
where we introduced the phase space function (sometimes called
\textquotedblleft nesting function\textquotedblright )%
\begin{equation}
\tilde{N}_{\mathbf{q}}(0)=\frac{1}{N(0)}\sum_{\mathbf{k}}\delta (\varepsilon
_{\mathbf{k}})\delta (\varepsilon _{\mathbf{k+q}}).  \label{jdos}
\end{equation}%
{For a 2D cylindrical Fermi surface we have}

\begin{equation}
\tilde{N}_{\mathbf{q}}(0)=\frac{\theta (1-Q)}{4\pi \varepsilon _{F}Q\sqrt{%
1-Q^{2}}},  \label{jdos2D}
\end{equation}%
{which diverges at }$Q\rightarrow 0${ and }$Q\rightarrow 1$%
. Following Ref. \cite{allen} we can introduce the
\textquotedblleft mode $\lambda $\textquotedblright\ \textit{via} the
expression $\lambda =\sum_{\mathbf{q}}\lambda _{\mathbf{q}}$. Then%
\[
\lambda _{\mathbf{q}}=\tilde{N}_{\mathbf{q}}(0)\frac{\left\vert
g^{0}\right\vert ^{2}}{M}\frac{1}{\Gamma _{\mathbf{q}}^{2}+\omega _{\mathbf{q%
}}^{2}}. 
\]%
In the weak-damping approximation for we recover the standard formula%
\begin{equation}
\lambda _{\mathbf{q}}^{app}=\tilde{N}_{\mathbf{q}}(0)\frac{\left\vert
g^{0}\right\vert ^{2}}{M}/\omega _{\mathbf{q}}^{2},  \label{lam0}
\end{equation}%
but for\textbf{\ }$\Gamma _{\mathbf{q}}\gg \omega _{\mathbf{q}}$\textbf{\ }%
the contribution of strongly damped phonons to total\textbf{\ }$\lambda _{%
\mathbf{q}}$\textbf{\ }is suppressed.

The result (\ref{lamg}) we can get also if we introduce, according to Eq. %
\ref{D}, a generalized Eliashberg function%
\begin{eqnarray}
\alpha _{\Gamma }^{2}(\omega )F(\omega ) &=&\frac{1}{2\pi MN(0)}\sum_{%
\mathbf{k,q}}\frac{\delta (\varepsilon _{\mathbf{k}})\left\vert
g^{0}\right\vert ^{2}\delta (\varepsilon _{\mathbf{k+q}})}{2\omega _{\mathbf{%
q}}}\frac{1}{2}\left[ \frac{\Gamma _{\mathbf{q}}}{(\omega -\omega _{\mathbf{q%
}})^{2}+\Gamma _{\mathbf{q}}^{2}}-\frac{\Gamma _{\mathbf{q}}}{(\omega
+\omega _{\mathbf{q}})^{2}+\Gamma _{\mathbf{q}}^{2}}\right] =  \nonumber
\label{eliG} \\
&&\frac{1}{2\pi M}\sum_{\mathbf{q}}\frac{\tilde{N}_{\mathbf{q}}(0)\left\vert
g^{0}\right\vert ^{2}}{2\omega _{\mathbf{q}}}\frac{1}{2}\left[ \frac{\Gamma
_{\mathbf{q}}}{(\omega -\omega _{\mathbf{q}})^{2}+\Gamma _{\mathbf{q}}^{2}}-%
\frac{\Gamma _{\mathbf{q}}}{(\omega +\omega _{\mathbf{q}})^{2}+\Gamma _{%
\mathbf{q}}^{2}}\right] .
\end{eqnarray}%
The second term in this expression cancels out the nonphysical behavior at
low and high frequencies. Otherwise $\lambda _{\Gamma }$ would have be
divergent. Eqs. \ref{lamg},\ref{eliG} are general and valid for not only for
the 2D systems, where phase space factor (Eq. \ref{jdos}) is divergent.

For $\Gamma _{\mathbf{q}}\ll \omega _{\mathbf{q}}$ we have

\begin{equation}
\alpha _{app}^{2}(\omega )F(\omega )=\frac{1}{N(0)M}\sum_{\mathbf{k,q}}\frac{%
\delta (\varepsilon _{\mathbf{k}})\left\vert g^{0}\right\vert ^{2}\delta
(\varepsilon _{\mathbf{k+q}})}{2\omega _{\mathbf{q}}}\delta (\omega -\omega
_{\mathbf{q}})=\frac{1}{2\pi N(0)}\sum_{\mathbf{q}}\frac{\Gamma _{\mathbf{q}%
}^{app}}{\omega _{\mathbf{q}}}\delta (\omega -\omega _{\mathbf{q}}),
\label{eliash0}
\end{equation}%
where in the last equality we have used the approximate Eq. \ref{g1}. 
{Note that the Eq. \ref{eliash0} is a consequence
of that fact that the damping }$\Gamma _{\mathbf{q}}^{app}${, according to Eq.
\ref{g1}, can be expressed via the ``nesting function'' (Eq. \ref{jdos}) and both
are determined by the same function }${\rm Im}\Pi ^{\prime }(q,0)\equiv d%
{\rm Im}\Pi (q,\omega )/d\omega |_{\omega =0}${. In a general case
these functions can be different.}
This result without
using the pole approximation for the phonon Green
function can be trivially
obtained in the Matsubara formalism. In this case
it one does not need to
solve the Dyson equation. In the lowest order in
coupling for $T=0$ the
self-energy has a form

\[
\Sigma (i\varepsilon ,\mathbf{k}) =-i\int 
\frac{d(i\omega )}{2\pi }\sum_{%
\mathbf{k}^{\prime },\nu }\left\vert
g^{0}\right\vert ^{2}D_{M\nu }(i\omega
+i\varepsilon ,\mathbf{k,k}^{\prime
}) \frac{1}{i\omega -\varepsilon _{%
\mathbf{k}^{\prime }}-\Sigma (i\omega ,%
\mathbf{k}^{\prime })}, 
\]%
where 
\begin{equation}
D_{M\nu }(i\omega ,%
\mathbf{k,k}^{\prime })=(1/\pi )\int_{0}^{\infty }d\Omega 
\mathrm{Im}%
D(\Omega +i\delta ,\mathbf{k,k}^{\prime })\left[ \left( i\omega
-\Omega
\right) ^{-1}-\left( i\omega +\Omega \right) ^{-1}\right] .
\label{Dmats}
\end{equation}%
We can also average the self-energy over the Fermi surface $%
\Sigma
(i\varepsilon )=\sum_{\mathbf{k}}\delta (\varepsilon _{\mathbf{k}%
})\Sigma
(i\varepsilon ,\mathbf{k})/N(0).$

\[
\Sigma (i\varepsilon ) =%
\frac{1}{N(0)}\sum_{\mathbf{k}}\sum_{\mathbf{q}%
}\delta (\varepsilon _{%
\mathbf{k}})\left\vert g^{0}\right\vert ^{2}\delta
(\varepsilon _{\mathbf{%
k+q}}) \int_{-\infty }^{\infty }\frac{d\omega }{2\pi }%
D_{M}(\mathbf{q}%
,i\omega )\int_{-\infty }^{\infty }\frac{d\xi }{%
i(\varepsilon -\omega
)-\xi } 
\]

The integral $\int_{-\infty }^{\infty }\frac{d\xi }{%
i(\varepsilon -\omega
)-\xi }=-i2\pi \mathrm{sign}(\varepsilon -\omega )$
allows to calculate the 
\textit{physical} coupling constant $\lambda $ 
\begin{eqnarray*}
\lambda &=&-\frac{\partial \Sigma _{el}(i\varepsilon )}{%
\partial
i\varepsilon }|_{\varepsilon \rightarrow 0}=-\lim_{\varepsilon
\rightarrow 0}%
\frac{1}{N(0)}\sum_{\mathbf{k}}\sum_{\mathbf{q}}\delta
(\varepsilon _{%
\mathbf{k}})\left\vert g^{0}\right\vert ^{2}\delta
(\varepsilon _{\mathbf{k+q%
}}) \int \frac{d\omega }{2\pi }D(\mathbf{q}%
,i\omega )2\pi \delta
(\varepsilon -\omega ) \\
&=&-\frac{1}{N(0)}\sum_{%
\mathbf{k}}\sum_{\mathbf{q}}\delta (\varepsilon _{%
\mathbf{k}})\left\vert
g^{0}\right\vert ^{2}\delta (\varepsilon _{\mathbf{k+q%
}})D_{M}(\mathbf{q}%
,i0) =-\frac{1}{N(0)}\sum_{\mathbf{k}}\sum_{\mathbf{q}%
}\delta
(\varepsilon _{\mathbf{k}})\left\vert g^{0}\right\vert
^{2}\delta
(\varepsilon _{\mathbf{k+q}})\frac{1}{D_{M0}^{-1}(i0)-\Pi _{M}(%
\mathbf{q},i0)}.
\end{eqnarray*}%
\end{widetext}
On
the Matsubara axes, for 2D system, according to Eq. \ref{mats}, the phase
space factor vanishes for $q>2k_{F}$ and $\Pi _{M}(q\leq 2k_{F},i0)=-2\zeta
\omega _{0}^{2}$ is a constant (see Eqs.\ref{mats} and \ref{re_small}).
Using Eqs. \ref{D0},\ref{Dmats} we get $D_{M0}^{-1}(i0)=-M\omega _{0}^{2}$
and

\begin{equation}
\lambda =\zeta /(1-2\zeta ).  \label{lam}
\end{equation}%
In a 3D case $\Pi _{M}(\mathbf{q},i0)$ is a rather complicated function of $%
q $ and the Eq. \ref{lam} is only an approximation (as was probably firstly
mentioned by Fr\"{o}hlich \cite{froeh}). The physical meanings of the
coupling constants $\lambda $ and $\zeta $ that they are measures of the
renormalization of the phonon frequency from $\omega _{0}$ to $\omega _{%
\mathbf{q}}$ (\textit{cf.} a discussion for 3D systems in Ref.\cite{maks} ).

Turning back to the 2D case, according to Eq. \ref{lamg} we can neglect all
divergent contributions near $Q=0$ and $Q=1$. Elsewhere we can use Eq. \ref%
{lam0}.

One should keep in mind that the \textit{conventional coupling constant} is $%
\lambda =\zeta /(1-2\zeta )$. This parameter determines electronic
properties (Fermi velocities, $T_{c}$, etc.). The other parameter, $\zeta
=\lambda /(1+2\lambda )<1/2$, defines the observable phonon frequency, $%
\omega _{\mathbf{q}}=\omega _{0}\sqrt{1-2\zeta }$.

\section{Conclusions}

First of all, the standard well-known expression%
\begin{eqnarray*}
&&\alpha ^{2}(\omega )F(\omega )=\frac{1}{2\pi N(0)}\sum_{\mathbf{q}}\frac{%
\Gamma _{\mathbf{q}}}{\omega _{\mathbf{q}}}\delta (\omega -\omega _{\mathbf{q%
}}) \\
&\equiv &\frac{1}{N(0)M}\sum_{\mathbf{k}}\sum_{\mathbf{q}}\frac{\delta
(\varepsilon _{\mathbf{k}})\left\vert g^{0}\right\vert ^{2}\delta
(\varepsilon _{\mathbf{k+q}})}{2\omega _{\mathbf{q}}}\delta (\omega -\omega
_{\mathbf{q}}),
\end{eqnarray*}%
is valid only in the lowest order in the \textquotedblleft
bare\textquotedblright\ phonon linewidth, $\Gamma _{\mathbf{q}}$, which is
not an acceptable approximation {in case of strong Kohn
singularities, and particularly} for a cylindrical Fermi surface. $\Gamma _{%
\mathbf{q}}$ in this approximation is not the actual phonon line width; as
opposed to $\Gamma _{\mathbf{q}},$ determined by the oversimplified Eq. \ref%
{g1}, the real phonon linewidth does not diverge even for an ideally
cylindrical Fermi surface. Second, the renormalized $\lambda $ for a
cylindrical Fermi surface and parabolic bands does not depend on filling.

\begin{acknowledgments}
IIM would like to thank the Max Planck Society for hospitality during
his visit to the Max Planck Institute for Solid State Research, where part
of this work was done.
\end{acknowledgments}


\begin{thebibliography}{99}
\bibitem{MA} I.I. Mazin and V.P. Antropov, Physica \textbf{C385}, 49 (2003).

\bibitem{kong} Y. Kong, O. V. Dolgov, O. Jepsen and O. K. Andersen, Phys.
Rev. \textbf{B64}, 020501(R) (2001).

\bibitem{amy} A.Y. Liu, I. I. Mazin and J. Kortus, Phys. Rev. Let. \textbf{87%
}, 087005 (2001).

\bibitem{an} J.M. An, S.Y. Savrasov, H. Rosner, and W.E. Pickett, Phys. Rev.
B \textbf{66}, 220502(R) (2002); W.E. Pickett, J.M. An, \ H. Rosner, and
S.Y. Savrasov, Physica C \textbf{387}, 117 (2003)

\bibitem{lib} A.Y. Liu, I. I. Mazin. Phys. Rev. \textbf{B75}, 064510 (2007);
M. Calandra, A.N. Kolmogorov, and S. Curtarolo, Phys. Rev \textbf{B75},
144506 (2007).

\bibitem{footnote} {In publications one can sometimes find
the phonon Green functions which differ from ours in some factors.
This ambiguity can be traced down to the definitions of the phonon field operators (Eq.
\ref{u}). In this case the corresponding factors appear in the
electron-ion matrix element (Eq. \ref{matr}). But all physical quantities as EPI
coupling constant, Eliashberg functions etc. are  determined by
the unique combination }$\left\vert g^{0}\right\vert ^{2}D${ that
is independent on definitions.}

\bibitem{migdal} A.B. Migdal, Sov. Phys.-JETP, \textbf{7}, 996 (1958)

\bibitem{allen} P.B. Allen in : \textit{Dynamical properties of solids},
eds. G.K. Horton and A.A. Maradudin, v. 3, ( North Holland, 1980), p. 157.

\bibitem{allen2} P.B. Allen, Phys. Rev. B \textbf{6}, 2577 (1972)

\bibitem{cal-maur} M. Calandra, and F. Maury, Phys. Rev. B \textbf{71},
064501 (2005)

\bibitem{stern} F. Stern, Phys. Rev. Letts., \textbf{18}, 546 (1967)

\bibitem{ando} T. Ando, A.B. Fowler, and F. Stern, Rev. Mod. Phys., \textbf{%
54}, 445 (1982)

\bibitem{zhang} Y. Zhang, V.M. Yakovenko, and S. Das Sarma, Phys. Rev. 
\textbf{B71},115105 (2005)

\bibitem{IT} A. Isihara and T. Toyoda, Z. Phys. \textbf{B23}, 389 (1976); 
{A. Isichara, in Solid State Physics, ed. by H. Erenreich, F. Zeitz,
and D. Turnbull, (Academic, N.Y., 1989), v. 42, p.271}

\bibitem{froeh} H. Fr\"{o}hlich, Phys. Rev. \textbf{79}, 845 (1950)

\bibitem{mahan} G.D. Mahan, \textit{Many-Particle Physics}, Plenum Press,
New York, 1981

\bibitem{am} P.B. Allen and B. Mitrovi\'{c}, in \textit{Solid State Physics}%
, ed. by H. Erenreich, F. Zeitz, and D. Turnbull, (Academic, N.Y., 1982), v. 
\textbf{37}, p.1

\bibitem{maks} E.G. Maksimov and D.I. Khomskii, in \textit{High-Temperature
Superconductivity} , ed. by V.L. Ginzburg and D.A. Kirzhnits, Consultants
Bureau, New York, 1982, Chap. 3.
\end{thebibliography}
\end{document}